\documentclass[symmetry,article,accept,moreauthors,pdflatex,10pt,a4paper]{Definitions/mdpi}

\usepackage{graphicx}
\usepackage{graphicx}
\usepackage{subfig}
\usepackage{caption}
\usepackage{amsmath}
\usepackage{amssymb}

\def\be{\begin{equation}}
\def\ee{\end{equation}}
\def\ba{\begin{eqnarray}}
\def\ea{\end{eqnarray}}

\def\nn{\nonumber}

\def\a{\mathbf{a}}
\def\b{\mathbf{b}}

\newcommand\mba{\mathbf{a}}
\newcommand\mbb{\mathbf{b}}
\newcommand\bee{\begin{equation*}}
\newcommand\eee{\end{equation*}}
\def\be{\begin{equation}}
\def\ee{\end{equation}}
\def\ba{\begin{eqnarray}}
\def\ea{\end{eqnarray}}

\def\nn{\nonumber}
\newcommand\dphi{\dot{\phi}}
\newcommand\eb{b_0 e^{-\alpha\phi}}

\def\a{\mathbf{a}}
\def\b{\mathbf{b}}

\firstpage{1} 
\pubvolume{xx}
\issuenum{1}
\articlenumber{5}
\pubyear{2019}
\copyrightyear{2019}
\history{Received: 08 October 2019; Accepted: 06 November 2019; Published: date}

\Title{Cosmological Solutions from a Multi-Measure Model with Inflaton Field}

\Author{Denitsa Staicova *$^{,\dagger}$\orcidA{} and Michail Stoilov $^{\dagger}$ }

\AuthorNames{Denitsa Staicova and Michail Stoilov}

\address [1]{%
  Institute for Nuclear Research and Nuclear Energy, Bulgarian Academy of Sciences, Tsarigradsko shosse 72, Sofia 1784, Bulgaria; mstoilov@inrne.bas.bg}

\corres{Correspondence: dstaicova@inrne.bas.bg}

\firstnote{These authors contributed equally to this work.}

\abstract{
In a recent work, we demonstrated that a modified gravity model in which a scalar ``darkon'' field is coupled to both the standard Riemannian metric and to another non-Riemannian volume form is compatible with observational data from Supernovae Type Ia. 
Here, we investigate a more complicated model with an additional ``inflaton'' scalar field. 
We demonstrate numerically that the model can qualitatively reproduce the Universe inflation epoch, matter
dominated epoch, and present accelerating expansion in a seamless way. 
We show that such solutions occur only when the model parameters are within a very particular range.
The main numerical problem we are faced with is reproducing the extremely small time of the inflation epoch.
Here, we present how the variation of some parameters affects this time. 
 }

\keyword{multi-measure model; inflation; dark energy and dark matter}

\begin{document}

\section{Introduction}
The observational data on the evolution of the Universe give very strong indications of the domination of different components of the energy density during different epochs. While the 
quantitative interpretation is model dependent, qualitatively, we know that the Universe has passed through different stages: the radiation dominated era, the matter dominated era, and the current, dark energy 
dominated era. Those epochs can be described in the frame of the $\Lambda-CDM$  model as different components in the energy density of the Universe. To resolve a number of cosmological 
problems, however, one needs to introduce an additional exponential expansion in the beginning of the Universe: inflation. While the initial inflation seems to solve the problems in front of 
the theory (namely the horizon problem, the flatness problem, the missing monopoles problem, and the large structure formation problem \cite{cosmo0}), its theoretical description is ambiguous. Numerous models have 
been proposed to describe inflation: chaotic inflation (one scalar field rolling in a potential), multi-field theories (more than one scalar field), modified gravity (f(R), Brans--Dicke, etc.); 
for a review, see \cite{Linde, Debono}. 

The Weyl (local conformal) invariance is believed to play an important role in quantum gravity, in particular with respect to solving the cosmology problem. This is because conformal symmetry forbids the existence of a cosmological constant, so the latter must appear as a dynamically generated constant of integration \cite{Oda}. Furthermore, in a conformally invariant theory of gravity, it is possible to create geodesically complete cosmologies, i.e., where one can integrate past singularities such as the one in the Big Bang or the ones found in the center of black holes (for other applications, see also \cite{Bars, 1904.04493, Edery}). While the most recent Planck results rule out the perfect scale invariance of the spectrum of primordial fluctuations in the CMB \cite{Planck2018}, they are still very close to scale invariance. Because of this, the Weyl invariance has inspired many works, using the Higgs mechanism, which is an example of spontaneous symmetry breaking of gauge symmetries, to produce simple Higgs non-minimally coupled to gravity inflation models (for example, see \cite{1807.02376}). 

Examples of theories inspired by the Weyl invariance are the multi-measure gravitational models developed by Guendelman, Nissimov, and Pacheva \cite{ref01,ref01_1, ref01_2, ref01_31, ref01_32, ref01_33, 1407.6281, 1408.5344, 1609.06915, 1507.08878, 1603.06231}. 
In these models, there are one or two scalar fields, coupled to more than one independent volume form. 
One of the volume forms is always the standard Riemannian volume form, and the others are dual to the derivative of the auxiliary third rank antisymmetric gauge field(s) (exact four-form). 
These models describe simultaneously both the dark energy that is dynamically generated (an integration constant in the simplest one scalar field model) and dark matter as a dust contribution to the energy momentum tensor.

In our previous work \cite{1610.08368, 1801.07133, 1906.08516}, we studied the cosmological aspect of a two measure model with a new scalar field (called darkon).
The model is capable of fitting the data of Supernova Type Ia and, moreover, of producing two families of parameters for which the model corresponds to the observational data. This model, however, was able to explain only the late-time evolution of the Universe, i.e., the matter dominated epoch and current accelerated expansion.

Here, we present our numerical results on the inflationary model proposed in \cite{1408.5344, 1609.06915}.
This is a two scalar field model (the scalars are called darkon and inflaton) with a global Weyl symmetry. 
In the model, the dynamics of the Universe is governed by the movement of the inflaton field over an effective step-like potential with two infinite plateaus connected with a steep slope. 
According to \cite{1408.5344, 1609.06915}, the left, higher plateau corresponds to the inflationary Universe, while the right, lower plateau to the current expanding Universe, and the inflaton is freely moving from the left to the right. Our results \cite{1906.08516} have shown that due to the existence of a friction term, the inflaton cannot move freely, so it has to start its motion from the slope of the potential. 
Here, we present our progress in finding a proper description of the Universe's evolution and how the variations of some of the parameters affect the time at which the inflation ends. 

\section{Overview of the Multi-Measures Model}
A multi-measure model aimed at describing the evolution of the Universe was developed in \cite{1408.5344, 1609.06915}. It features two scalar fields: an inflaton $\phi$ and darkon $u$. 
Here, we mention only the points directly related to our numerical work. 
The action of the model is: 

\be
S= S_{darkon}+S_{inflaton}
\label{totA}
\ee
where $S_{darkon}$ and $S_{inflaton}$ are the darkon and inflaton actions, respectively. 

In $S_{darkon}$, {two} independent measures $\sqrt{-g}$ and $\Phi(C)$ are coupled to the darkon Lagrangian $L$: 

\be
 S_{darkon}=\int{d^4x(\sqrt{-g}+\Phi(C))L}.
\label{darA}
\ee

Here, $\Phi(C)$ is an integration measure density dual to the field strengths of an auxiliary three-index antisymmetric tensor gauge field $C_{\nu\kappa\lambda}$, namely:

\be
\Phi(C)=\frac{1}{3\!}\epsilon^{\mu\nu\kappa\lambda}\partial_\mu C_{\nu\kappa\lambda},
 \label{TMG}
\ee

and:

\be
 L=-\frac{1}{2}g^{\mu\nu}\partial_\mu u\partial_\nu u -W(u). \label{drkL}
\ee 
 
We have {three} additional non-Riemannian measures in the inflaton action:

\be
S_{inflaton}=\int d^4x \Phi(A)(R+L^{(1)})+\int d^4x\Phi(B)\left(L^{(2)}+\frac{\Phi(H)}{\sqrt{-g}}\right).
\label{inflA}
\ee

Here, $A, B$, and $H$ are three additional auxiliary three-index antisymmetric tensor gauge fields, and $\Phi(Z)$, for $Z=A,B,H$, is of the same form as Equation (\ref{TMG}).
Note that $R$ is in the Palatini formalism, so we have $\Gamma^\lambda_{\mu\nu}$ as independent fields as well, and: 

\begin{align}
& L^{(1)}=-\frac{1}{2}g^{\mu\nu}\partial_\mu\phi\partial_\nu \phi- V(\phi),\; V(\phi)=f_1 e^{-\alpha \phi} \\
&L^{(2)}=-\frac{b_0}{2}e^{-\alpha\phi} g^{\mu\nu}\partial_\mu\phi\partial_\nu \phi + U(\phi),\; U(\phi)=f_2 e^{-2\alpha \phi}
\end{align}
\normalsize

The gauge fields $Z=H,A,B,C$ are pretty much like electromagnetic potentials;
however, they are not vectors (more precisely, they are not one-forms), but rank-three antisymmetric tensors (three-forms). Their field strengths $\mathcal{F}(Z)$ that determine the measures $\Phi(Z)$ are $\mathcal{F}(Z)=\partial_{[{\mu}} Z_{\nu \rho \gamma]}$, where $[...]$ denotes antisymmetrization of the indexes in the brackets. Note that $\mathcal{F}(Z)$ are tensors (four-forms) because of antisymmetrization (on the forms, the exterior derivative and antisymmetrized covariant derivative are equivalent). The gauge transformation under which the model action is invariant is $Z_{\mu\nu\rho}\rightarrow Z_{\mu\nu\rho} + \partial_{[\mu}z_{\nu\rho]}$, where $z=\tilde{A},\tilde{B},\tilde{C},\tilde{H}$ are four arbitrary rank-two antisymmetric tensors (two-forms).

A very interesting and unusual property of the action Equation (\ref{darA}) is that the dynamics does not depend on the particular form of the potential $W(u)$.
If we choose it to be: 

\be 
W(u)\sim u^4
\ee
then the action Equation (\ref{totA}) is invariant under the following global Weyl scale transformation:

\ba 
&& g_{\mu\nu}\rightarrow \lambda g_{\mu\nu},\;\;
\Gamma^\rho_{\mu\nu}\rightarrow\Gamma^\rho_{\mu\nu},\;\;
\phi\rightarrow\phi+\frac{1}{\alpha}\ln\lambda,\;\;
u\rightarrow \frac{1}{\sqrt{\lambda}} u,
\label{winv}\\
&&A_{\mu\nu\rho}\rightarrow \lambda A_{\mu\nu\rho},\;\;
B_{\mu\nu\rho}\rightarrow \lambda^2 B_{\mu\nu\rho},\;\;
C_{\mu\nu\rho}\rightarrow \lambda^2 C_{\mu\nu\rho},\;\;
H_{\mu\nu\rho}\rightarrow H_{\mu\nu\rho}
\nonumber
\ea

The variation of $S$ with respect to the auxiliary fields $A, B, C$, and $H$ leads to four
 dynamically generated integration constants $M_0,\;M_1,\;M_2$, and $\chi_2$:

 \ba
 L&=&-2M_0, \nonumber\\
 R+L^{(1)}&=&-M_1, \nonumber\\
 L^{(2)}+\frac{\Phi(H)}{\sqrt{-g}}&=&-M_2,\nonumber\\
 \frac{\Phi(B)}{\sqrt{-g}}&=&\chi_2.
 \label{const} 
\ea

Note that only $\chi_2$ is dimensionless, while $M_0, M_1$ and $M_2$ have dimension $\mathrm{mass}^4$ (so do $f_1$ and $f_2$).
The appearance of dimensionful constants as solutions of equations of motion signals dynamical spontaneous breaking of the symmetry Equation (\ref{winv}).

Equations (\ref{const}) together with the equation following from the variation with respect to $\Gamma$ can be used to eliminate the auxiliary fields in the model, leaving an effective Lagrangian depending only on $g_{\mu\nu}$ and the inflaton scalar field. Explicitly, for the Weyl rescaled metric $\tilde{g}_{\mu\nu}$ and the redefined darkon scalar field $\tilde{u}$:
\ba 
\tilde{g}_{\mu\nu}=\frac{\Phi(A)}{\sqrt{-g}} g_{\mu\nu}&\\
\frac{\partial\tilde{u}}{\partial u}=(W(u)-2M_0)^{-\frac{1}{2}},&
\ea
we get the standard action $S^{(eff)}=\int{d^4x \;\sqrt{-\tilde{g}}(\tilde{R}+L^{(eff)})}\label{sef}$ for the effective Lagrangian: 

\be
L^{(eff)}=\tilde{X}-\tilde{Y}(V+M_1-\chi_2\eb\tilde{X})+\tilde{Y}^2(\chi_2(U+M_2)-2M_0).
\label{l_new}
\ee

Here, $\tilde{Y}=-\frac{1}{2}\tilde{g}^{\mu\nu}\partial_\mu\tilde{u} \partial_\nu\tilde{u} ,
\tilde{X}=-\frac{1}{2}\tilde{g}^{\mu\nu}\partial_\mu\phi \partial_\nu\phi 
$ are the respective kinetic terms for the new metric. This Lagrangian is non-linear with respect to both scalar fields' kinetic terms and thus can be classified as a generalized k
-essence type. 
For this, the Einstein equations correspond to the equations of motion derived for the original metric. Its step-by-step derivation can be found in \cite{1906.08516}.

In the Friedman--Lemaitre--Robertson--Walker space-time metric, the effective equations of motion are: 

\ba
v^3+3\mba v+2\mbb&=&0 \label{sys1}\\
\dot{a}(t)-\sqrt{\frac{\rho}{6}}a(t)&=&0 \label{sys2}\\
\frac{d}{dt}\left( a(t)^3\dphi(1+\frac{\chi_2}{2}\eb v^2) \right)+\hspace{25mm}&&\nonumber\\
a(t)^3 (\alpha\frac{\dphi^2}{4}\chi_2\eb v^2+\frac{1}{2}V_\phi v^2-\chi_2 U_\phi\frac{v^4}{4})&=&0\label{sys3}
\ea

The dot over the fields indicates the time derivative.
The algebraic Equation (\ref{sys1}) is a remnant of the darkon field equation where
the parameters are:
 $$\mba_{}=-\frac{1}{3}\frac{V(\phi)+M_1-\frac{1}{2}\chi_2 b_0 e^{-\alpha\phi}\dot{\phi}^2}{\chi_2(U(\phi)+M_2)-2M_0}, \mbb_{}=-\frac{p_u}{2a(t)^3(\chi_2(U(\phi)+M_2)-2M_0)}$$ 
and $p_u$ is an integration constant.
Equation (\ref{sys2}) is the first Friedman equation where 
$a(t)$ is the metric scaling function, and the energy density is:
 $$\rho=\frac{1}{2}\dphi^2 (1+\frac{3}{4}\chi_2 b_0 e^{-\alpha\phi} v^2)+\frac{v^2}{4} (V+M_1)+
 \frac{3 p_u v}{4a(t)^3}. $$
 
Equation (\ref{sys3}) is the effective inflaton equation of motion, where
$U_\phi=\frac{\partial U}{\partial \phi}$ and
 $V_\phi=\frac{\partial V}{\partial \phi}$.
 We can write down
the second Friedman equation as well: 

\begin{equation}
\ddot{a}(t)=-\frac{1}{12}(\rho+3p)a(t),\label{dda} 
\end{equation}
where $p=\frac{1}{2}\dphi^2 (1+\frac{1}{4}\chi_2\eb v^2)-\frac{1}{4}v^2(V+M_1)+p_u v/(4 a(t)^3)$.
However, Equation (\ref{dda}) is not independent, but is connected through the equation of state Equation (\ref{sys1}) to Equation (\ref{sys2}).


\section{Numerical Results}
 Before starting with the numerical analysis, it is useful to have some estimates of the asymptotic values of the energy density $\rho$ and pressure $p$ in the Universe to compare.

First, we can derive the following limits for the equation of state (EOS) $w=p / \rho$:

\ba
w & \xrightarrow[a(t)=0]{}& 1/3 \;\;\nn\\
w & \xrightarrow[a(t)\to \infty]{}& -1\label{asym2}
\ea

Second, the model possesses an ``effective potential'' depending only on the inflaton field $\phi$ of the~form:

\begin{equation}
U_{eff}(\phi)=\frac{(f_1 e^{-\alpha\phi}+M_1)^2}{4\chi_2 (f_2 e^{-2\alpha\phi}+M_2)-8 M_0}.
\end{equation}

For certain parameters, $U_{eff}(\phi)$ has a step-like form with a steep slope, connecting left (higher) and right (lower) plateaus. The length of the $\phi$-interval of the slope depends mainly on the parameter $\alpha$. 
The asymptotic of the left plateau is: 
$$U_{-}=U_{eff}\vert_{\phi\rightarrow -\infty}=\frac{f_1^2}{4\chi_2 f_2}$$
 and the asymptotic of the right plateau is:
$$U_{+}=U_{eff}\vert_{\phi\rightarrow +\infty}=\frac{M_1^2}{4\chi_2 M_2-8 M_0}.$$

This potential (in the regions where the kinetic energy of the inflaton can be neglected) can be connected to the effective cosmological constant in the model.
In this way, the model naturally has states with a (arbitrary) large and (arbitrary) small cosmological constant.
Note however that the effective potential does not bring the kinetic energy in standard form. For example, even in the limit of slow roll approximation (neglecting the terms $\sim\;\dot{\phi}^2,\;\; \dot{\phi}^3,\;\; \dot{\phi}^4$), the inflaton equation has the form: 

\be 
A\ddot{\phi}+3HA\dot{\phi}+U_{eff}'=0,
\label{eq_infl}
\ee 
where $A=1+b_0\,e^{-\alpha \phi}(V+M_1)/(2(U+M_2))$.
Nevertheless, we find the effective potential useful, because it naturally distinguishes three regions of $\phi$-values (left plateau, slope, right plateau) on which we observe three qualitatively different solutions of the system of Equations (\ref{sys1})--(\ref{sys3}), 
which enables us to track the movement of the inflaton correctly regardless of the validity of Equation (\ref{eq_infl}).

We demonstrated \cite{1906.08516} that in the Friedman--Lemaitre--Robertson--Walker space-time metric, when the scaling parameter $a(t)$ grows exponentially, there is a strong friction term in the scalar field equation of motion.
Therefore, in the far future, we have $\dot{\phi}(t)\to 0$ and a dynamically generated asymptotic cosmological constant, which for $\phi\rightarrow\infty$ is:

\be
\Lambda_{asymp}=\frac{U_+}{2}=\frac{M_1^2}{8(\chi_2 M_2-2M_0)}\label{coscon}.
\ee

We perform our calculations in units in which $c=1$, $G=1/16\pi$, and $t_u=1$, where $c$ is the speed of light, $G$ is Newton's constant, and $t_u$ is the present day age of the Universe.
Thus, our mass unit is equal to $1.62 \times 10^{59}M_{Pl}$ where $M_{Pl}$ is the Plank mass, the cosmological constant is $\Lambda\sim3.6$, and the Hubble constant is $H(1)\sim 1$.

The system of Equations (\ref{sys1})--(\ref{sys3}) has 12 free parameters, namely 
$\{\alpha,b_0, M_0, M_1, M_2, f_1,$ $f_2, p_u, \chi_2\}$.
Some of them are the theory parameters, and some are dynamically generated (integration constants). 
In addition, because Equation (\ref{sys2}) is a first order ordinary differential equation (ODE) and Equation~(\ref{sys3}) is a second order ODE, we have three initial conditions $\{a(0), \phi(0), \dot \phi (0)\}$.
Note that the differential Equation (\ref{sys2}) has a singular point in ${a}(0) = 0 $ due to the term $\sim 1/a(t)^3$ in $\rho$.  
Therefore, a natural replacement of the initial condition $a(0)=0$ is the normalization condition $a(1)=1$, where $t=1$ corresponds to the current moment.
Since these are plenty of parameters, we would like to restrict somehow their number and the region of variation, and this is the main objective of the present work.

It was assumed in \cite{1408.5344} that the left plateau corresponds to the pre-inflationary Universe (Planck times) and the right plateau to the current and future accelerated expansion, 
On this basis, the following estimated values have been advocated: $M_1=10^{-60}M_{Pl}^4$, $M_2=M_{Pl}^4$, $f_1\sim 10 ^{-8}M_{Pl}^4, f_2 \sim 10 ^{-8}M_{Pl}^4$. 
Implicitly, these estimates are based on the existence of a solution describing the inflaton moving from the left to right plateau.
We do not observe numerically such a solution. 
Moreover, because of the friction in the inflaton equation of motion, there are strong indications that such a solution does not exist.
Therefore, the only restriction to the model parameters from the effective potential is:
\be
\frac{f_1^2}{f_2}>>\frac{M_1^2}{M_2}.
\label{res1}
\ee

Our strategy to deal with the system of Equations (\ref{sys1})--(\ref{sys3}) is to solve algebraically Equation~(\ref{sys1}), to substitute its solution in Equations (\ref{sys2}) and (\ref{sys3}), and then to integrate the resulting system numerically. We do that using the Fehlberg fourth - 
fifth order Runge--Kutta method with degree four interpolation implemented in Maple.

Equation (\ref{sys1}), as a cubic equation, has three roots, of which at least one is always real. 
The roots can be written as:

\be
v_i=\left(\sqrt[3]{-1}\right)_i
\frac{\a}{\mathcal{A}}- 
\overline{\left(\sqrt[3]{-1}\right)}_i\mathcal{A}
;\;\; i=1,2,3
\label{cubS}
\ee
where $\mathcal{A}=\sqrt[3]{-\b+\sqrt{\a^3+\b^2}}$ and $\left(\sqrt[3]{-1}\right)_i$ is one of the three roots of
 $\sqrt[3]{-1}$ $\left(= -1, (1+i\sqrt{3})/2, (1-i\sqrt{3})/2\right)$.
We have to work with the real solution of Equation (\ref{sys1}). 
However, there is no globally defined smooth real solution of Equation (\ref{sys1}) in the plane $[\mba,\mbb]$.
In this work, we will work with 
 $\mbb<0$, so the root $v_1$ is real, and we use it as a solution of Equation (\ref{sys1}) in what follows.
In order to have negative $\mbb$, we usually (but not always) use negative $M_0$.
In fact, we expect that we can fix $M_0$ to an arbitrary value (provided $\mbb<0$) because we expect that there will be a family of solutions of the model, parametrized by $M_0$ and giving an equally good description of the Universe evolution.
Our logic is that first, we showed in \cite{1610.08368} that with a proper combination of other parameters in the purely darkon model, we have a nice fit of Supernovae Type Ia data for any $M_0$,
 and second, we expect that at the observed supernovae times, the inflaton field is settled down, leaving the Universe's dynamics to be governed by the darkon field only (with some constants determined by the asymptotic value of the inflaton).
 
In \cite{1906.08516}, we connected the observed cosmological constant $\Lambda \sim 3.6$ to $\Lambda_{asymp}$, and we used Equation (\ref{coscon}) to eliminate one of the parameters $M_0, M_1, M_2$, and $\chi_2$.
 However, in this way, we obtained a slightly larger Hubble constant because, first, the final value of $\phi$ is not infinite, but finite, and second, the movement of the inflaton is not stopped at $t=1$. 
Therefore, here, we use $\Lambda_{asymp}=1$, which gives $H(1)\sim 0.7$.

We observe that practically, the evolution does not depend on $\dot{\phi}(0)$; any initial inflaton velocity is almost immediately reduced to the one, dictated by the dynamics. Numerically, the velocity (up to $\dot{\phi}(0)\sim \pm10^4$) is corrected in the first step of the integration with minimal deviation of the other variables.
This gives us the possibility to use more ``physical'' initial conditions: 

\be \dot{\phi}(0)=0.\label{res3}\ee

\begin{figure}[H]\centering{
\includegraphics[width=5cm]{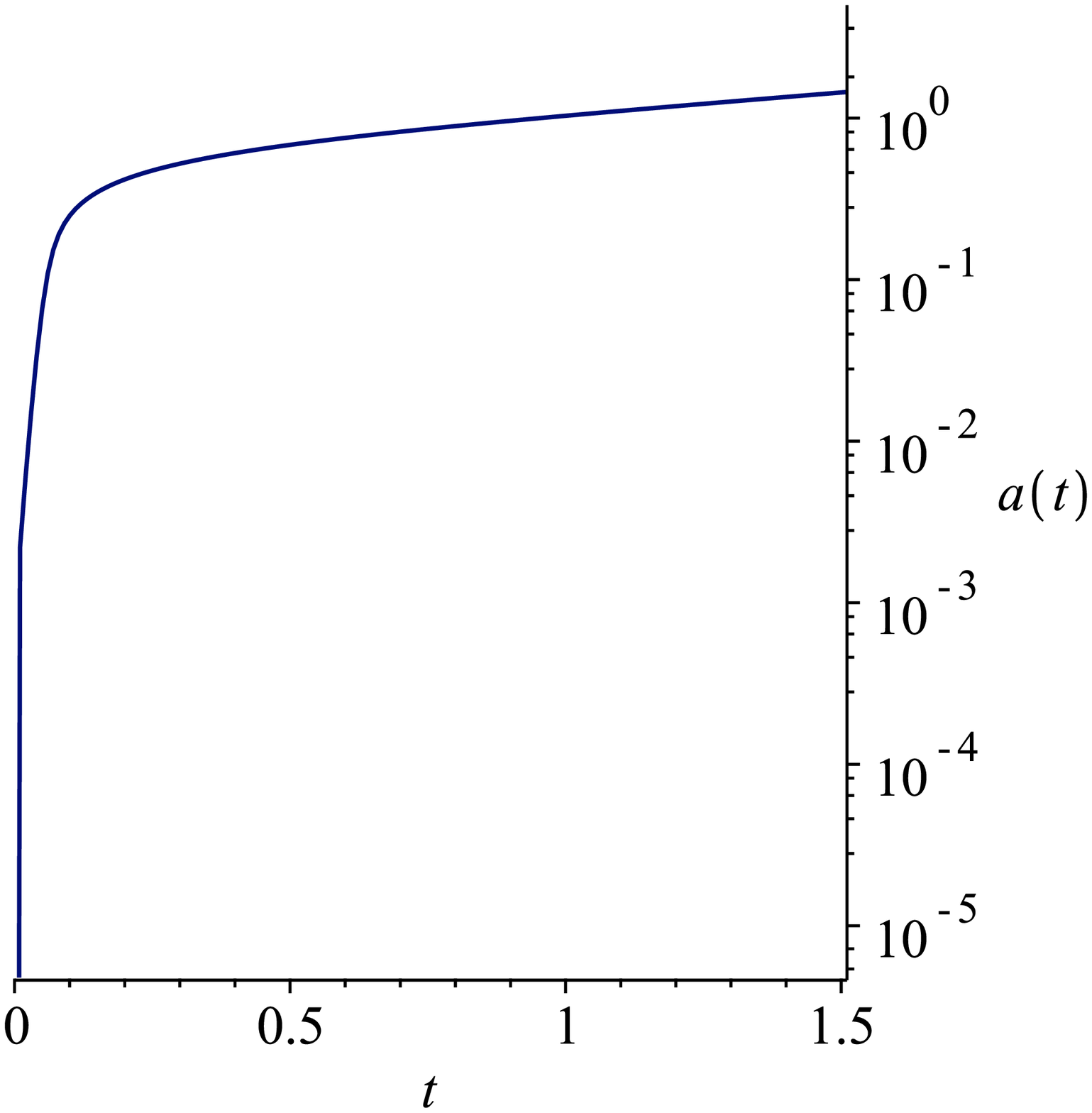} 
\includegraphics[width=5cm]{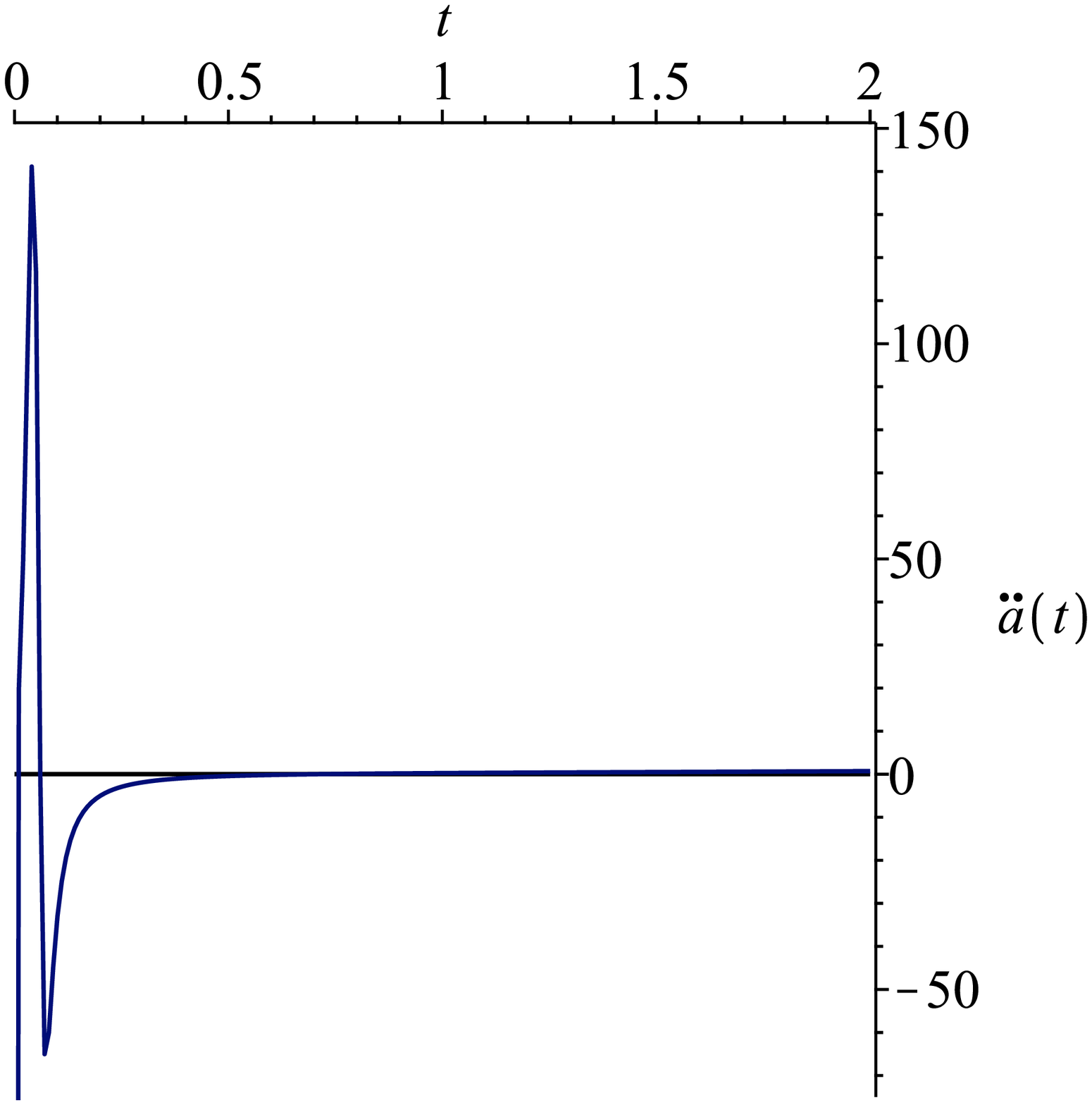}
\\
a \hskip 5cm b\\
\includegraphics[width=5cm]{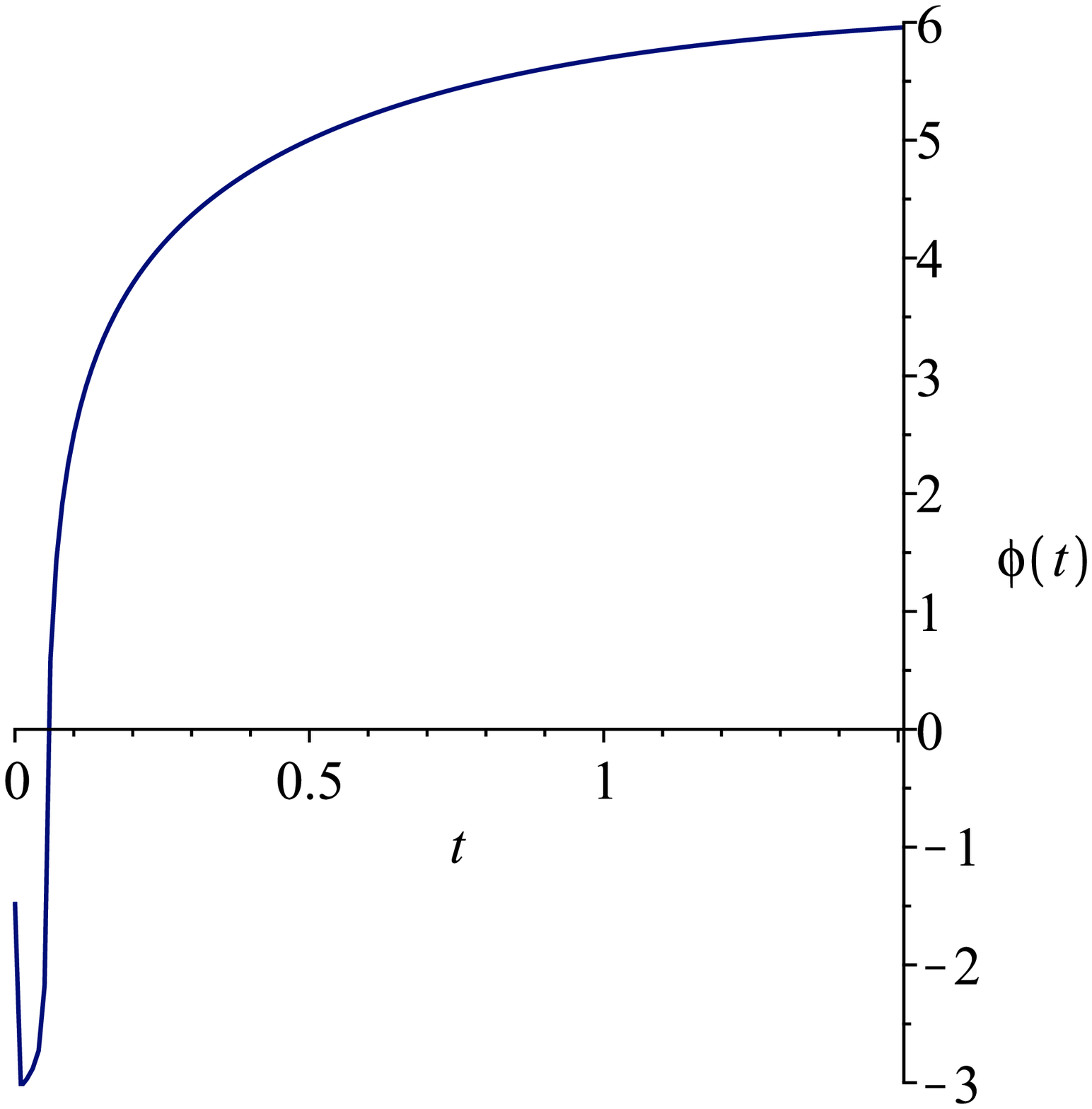} 
\includegraphics[width=5cm]{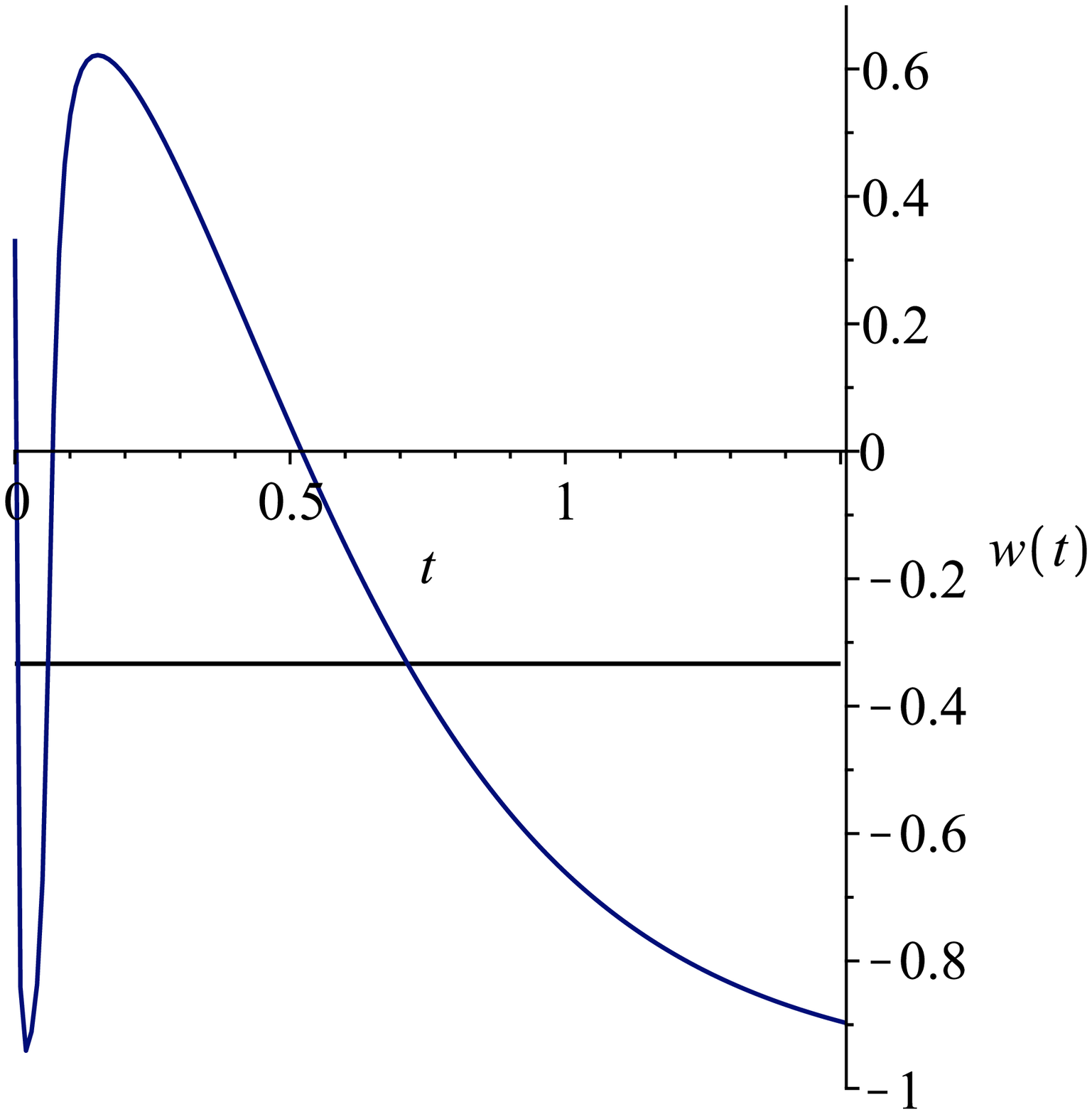}\\
c \hskip 5cm d

}
 \caption{The Universe's evolution for $(\chi_2=1,\;M_0=0.0001,\;M_1=0.08,\;M_2=0.001)$.
 The parameters $\{\alpha,b_0, p_u, f_1,f_2\}$ are 
 $\{1.1, 0.00119, 1.5\times10^{-7}, 20,10^{-3}\}$.
 On the panels are the evolution of: 
 (\textbf{a}) the scale factor $a(t)$, 
 (\textbf{b}) the second derivative of scale factor $\ddot{a}(t)$, 
 (\textbf{c}) the inflaton field $\phi(t)$, and 
 (\textbf{d}) the equation of state $w(t)=p(t)/\rho(t)$.}
 \label{Fig1}
 \end{figure}

\begin{figure}[H]\centering{
\includegraphics[width=5cm]{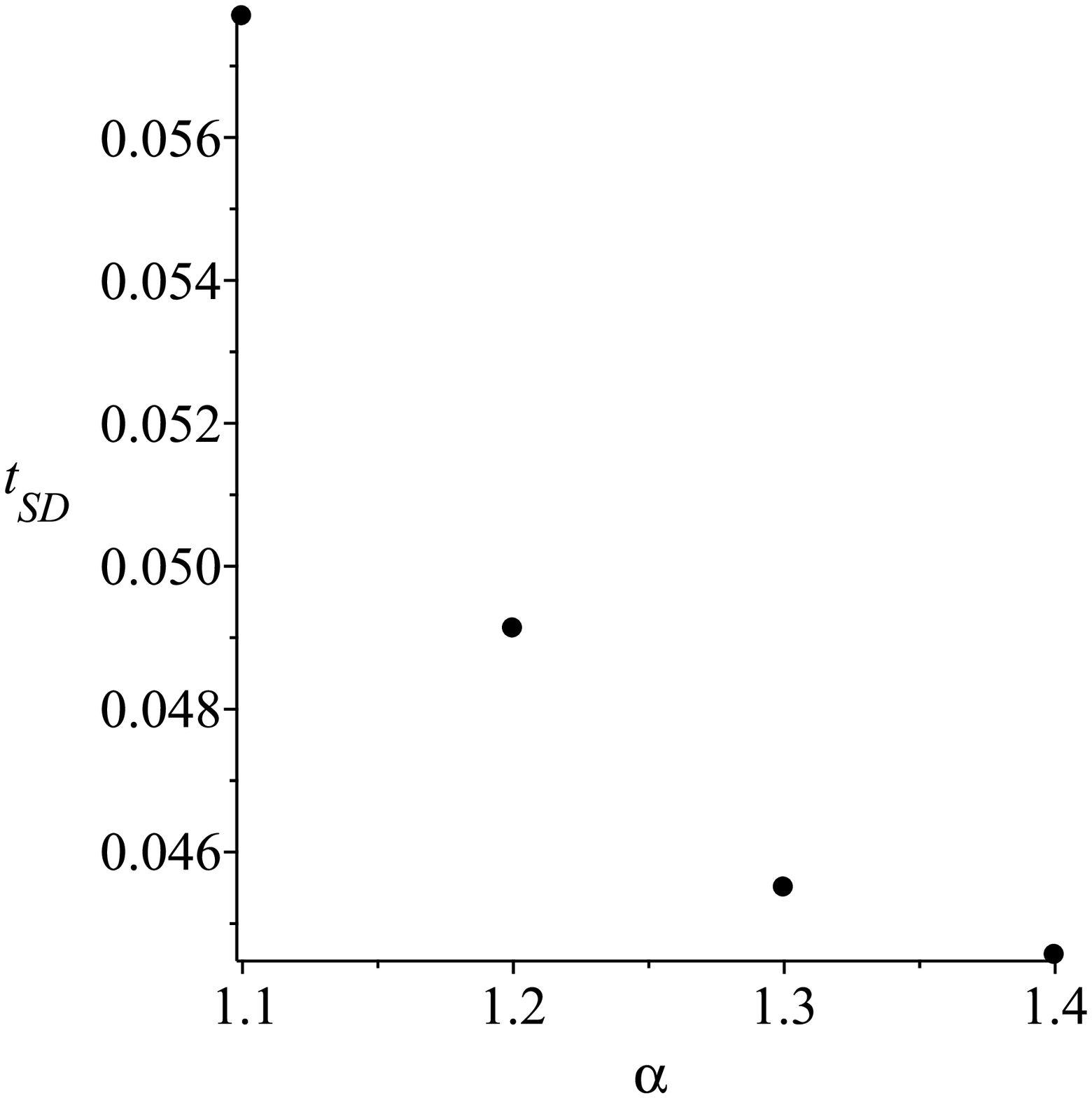} 
\includegraphics[width=5cm]{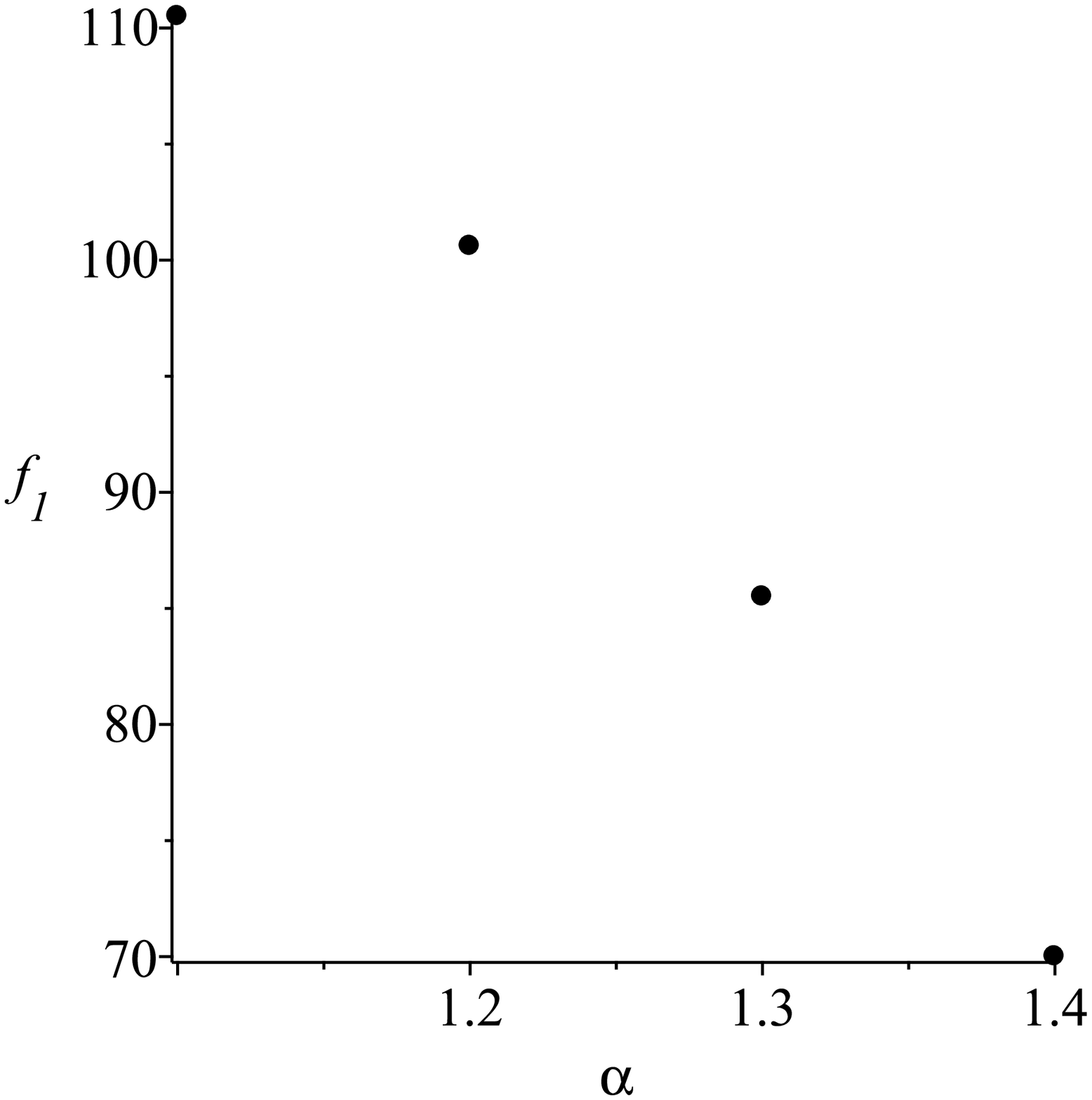}
\includegraphics[width=5cm]{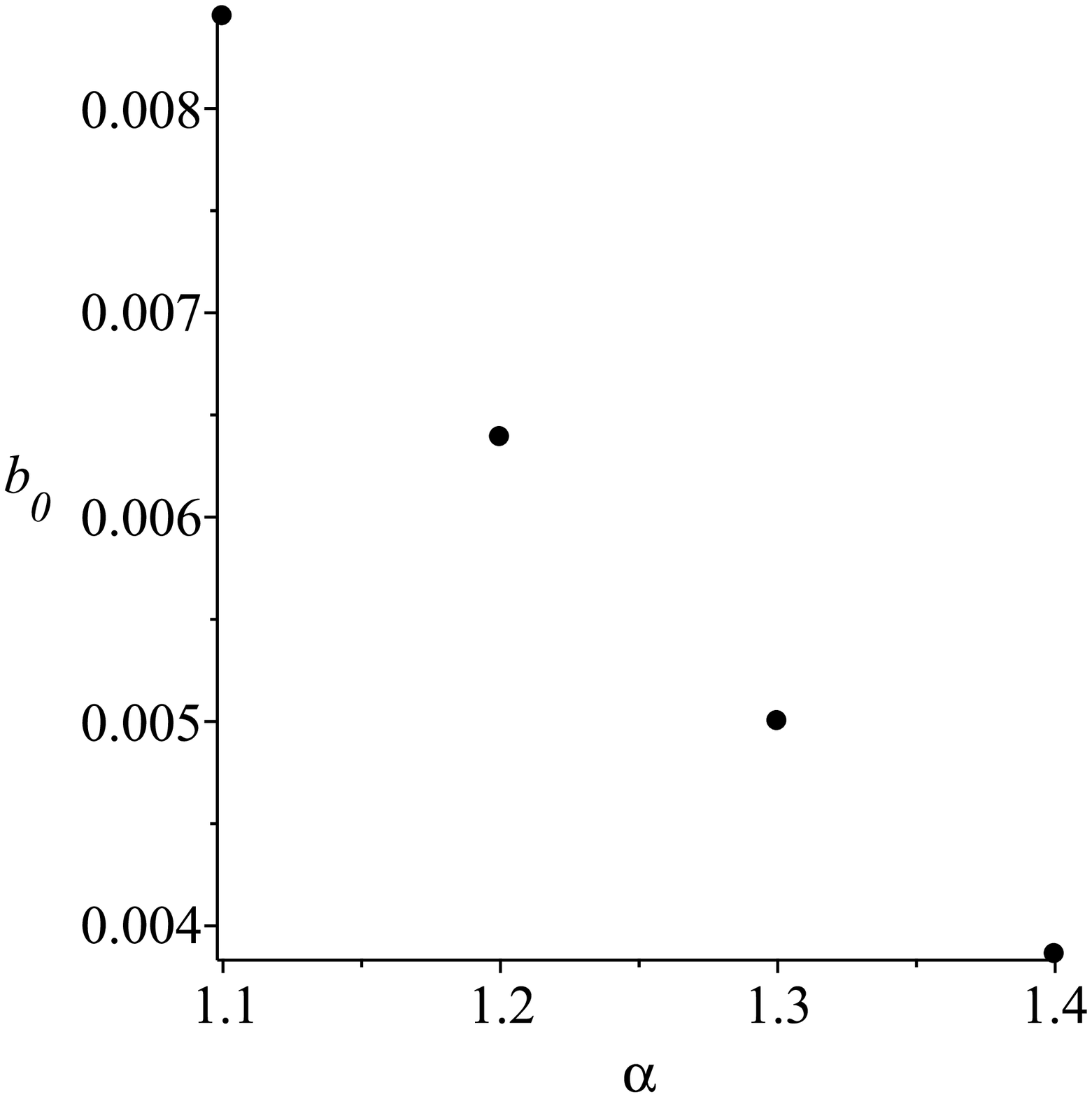}
\\
a \hskip 5cm b\hskip 5cm c
}
 \caption{(\textbf{a}) The beginning of the matter dominated epoch $t_{\mathrm{SD}}$ as a function of the parameter $\alpha$ for $\chi_2=1,\;M_0=-1,\;M_1=4,\;M_2=0.001, p_u=10^{-12},f_2=0.001$. (\textbf{b}) The $f_1$ parameters used. (\textbf{c}) The $b_0$ parameters used.}
 \label{Fig2}
 \end{figure}

\begin{figure}[H]\centering{
\includegraphics[width=5cm]{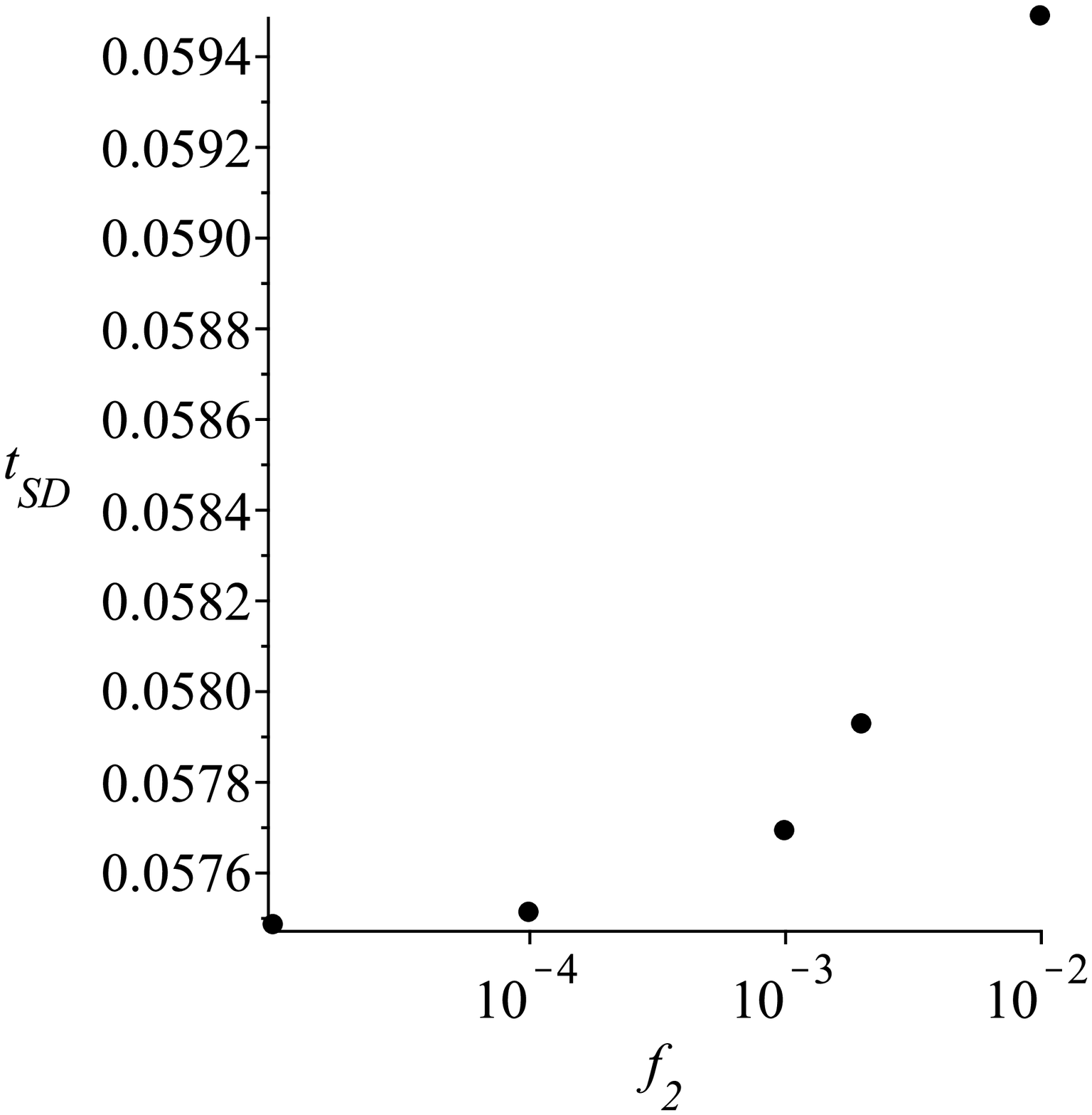} 
\includegraphics[width=5cm]{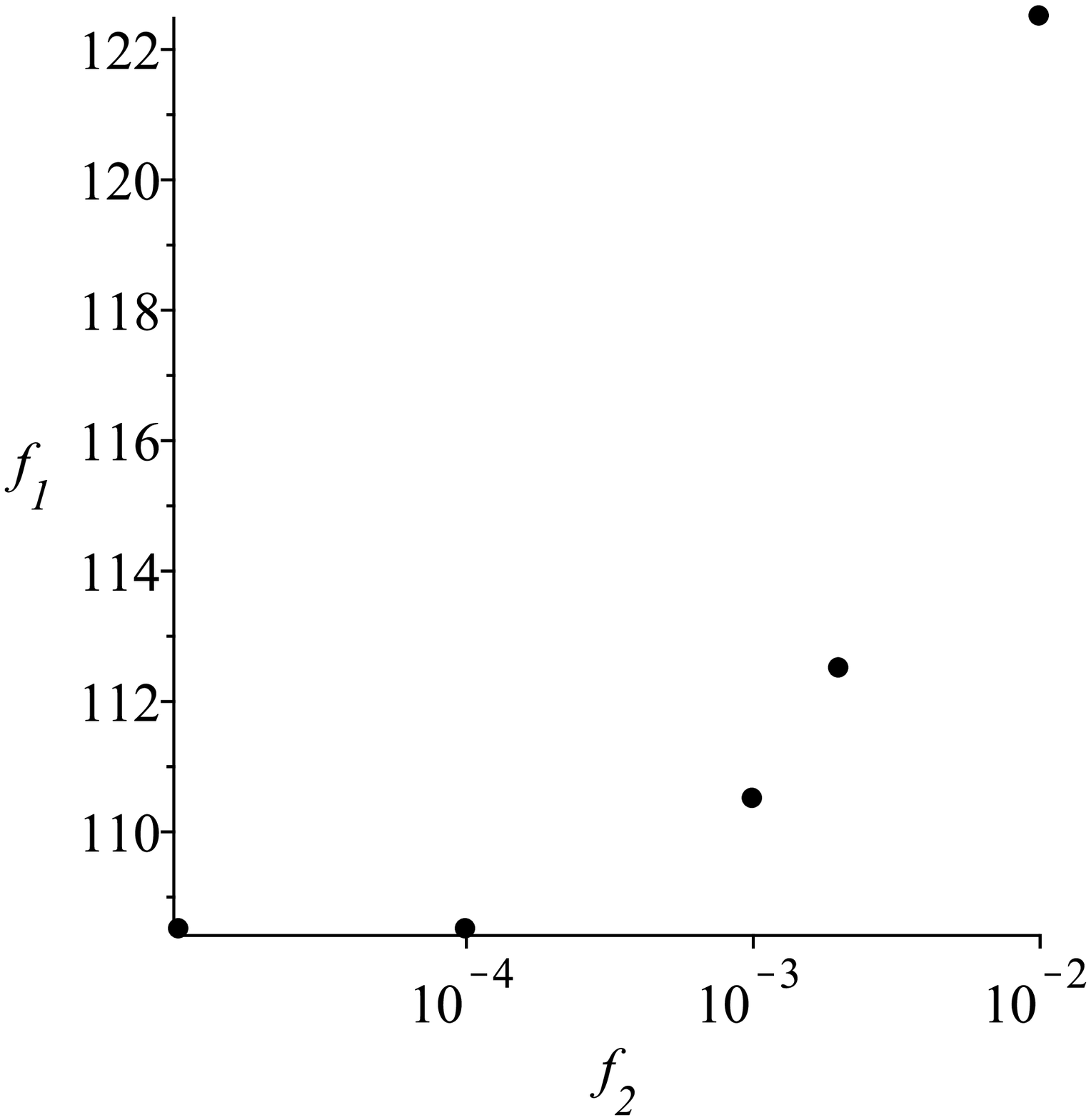}
\includegraphics[width=5cm]{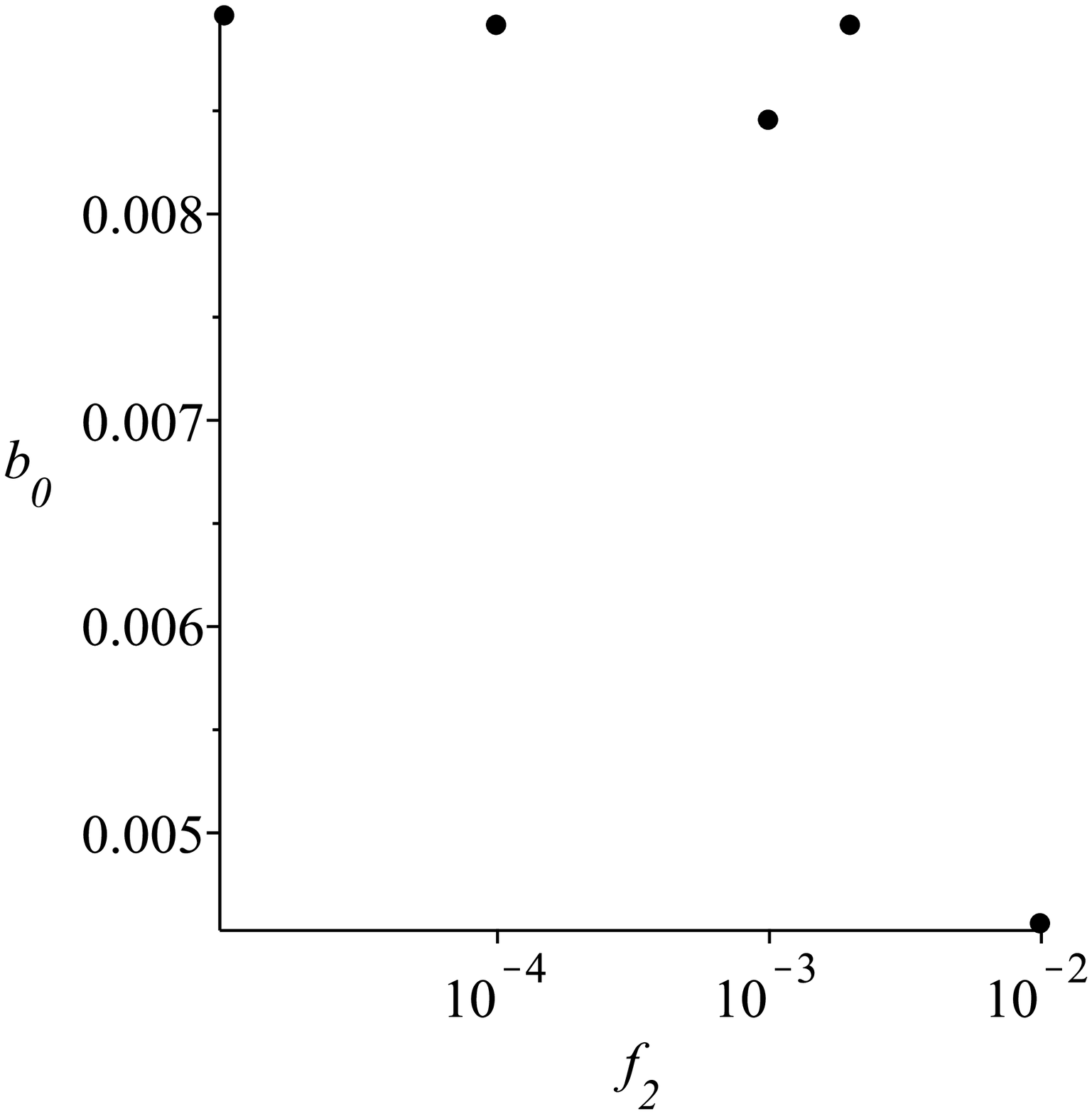}
\\
a \hskip 3cm b\hskip 3cm c
}
\caption{(\textbf{a}) The beginning of the matter dominated epoch $t_{\mathrm{SD}}$ as a function of the parameter $f_2$ for $\chi_2=1,\;M_0=-1,\;M_1=4,\;M_2=0.001, p_u=10^{-12}, \alpha=1.1$. (\textbf{b}) The $f_1$ parameters used. (\textbf{c}) The $b_0$ parameters used.}
 \label{Fig3}
 \end{figure}

\begin{figure}[H]\centering{
\includegraphics[width=5cm]{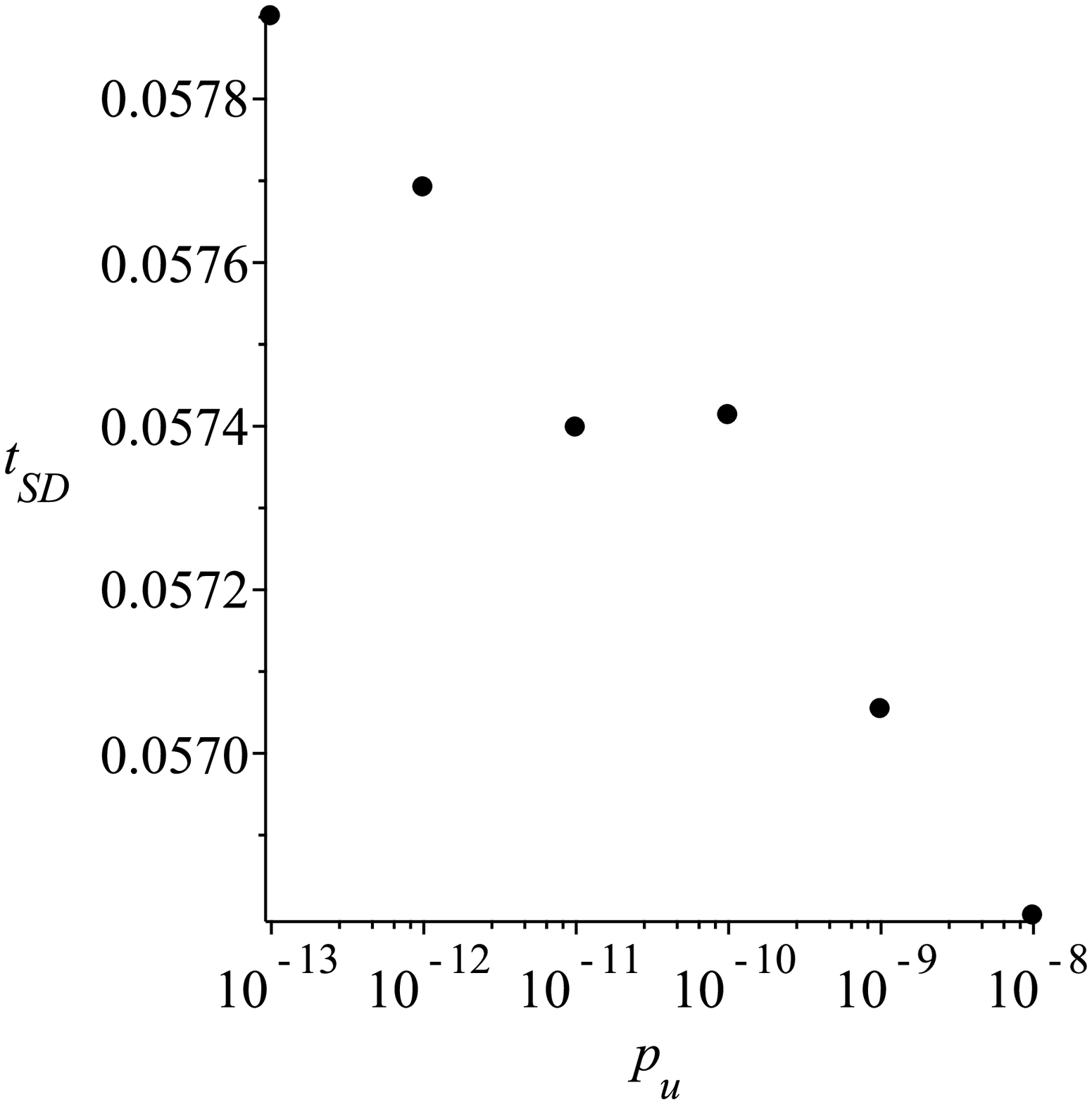} 
\includegraphics[width=5cm]{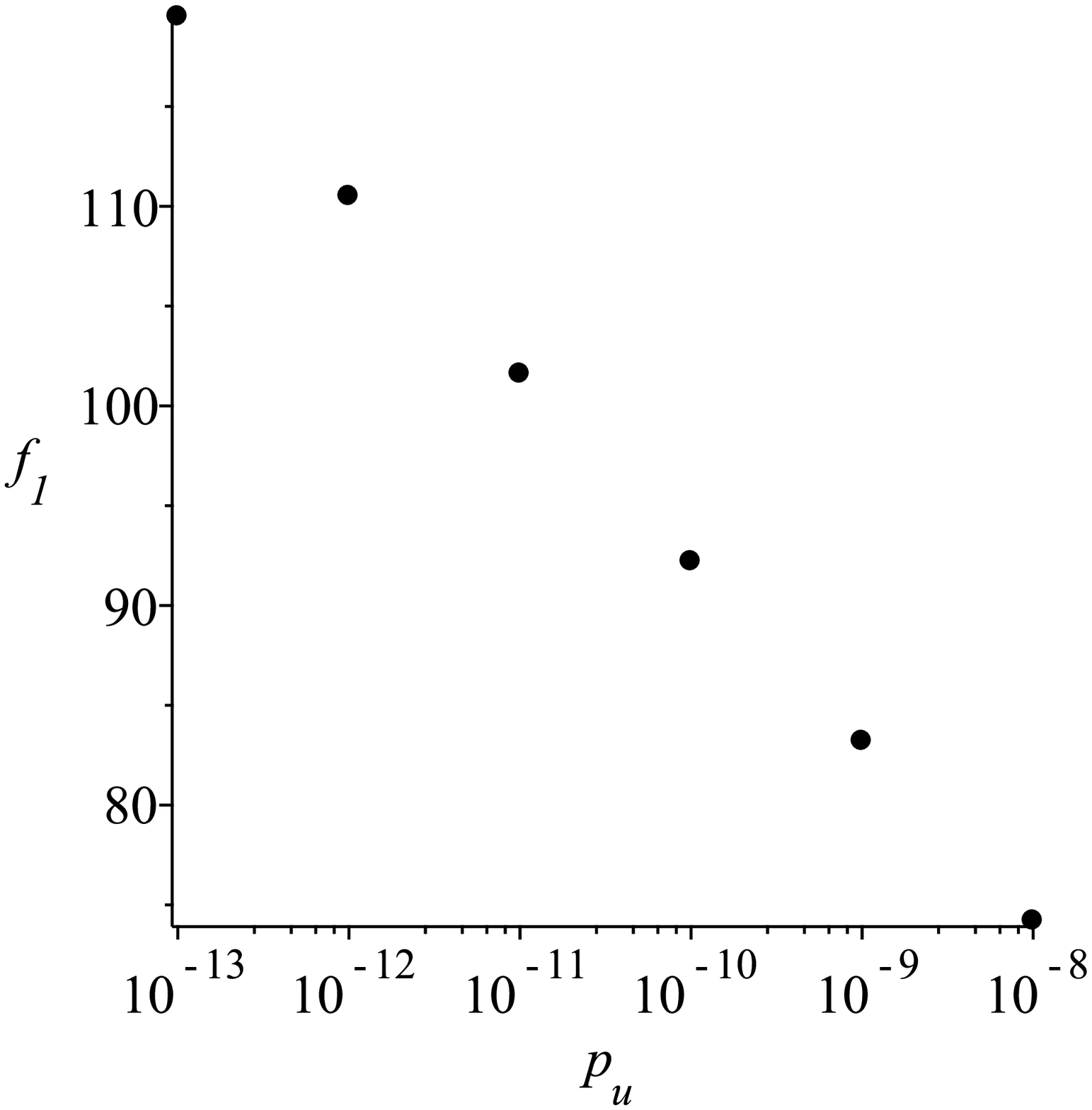}
\includegraphics[width=5cm]{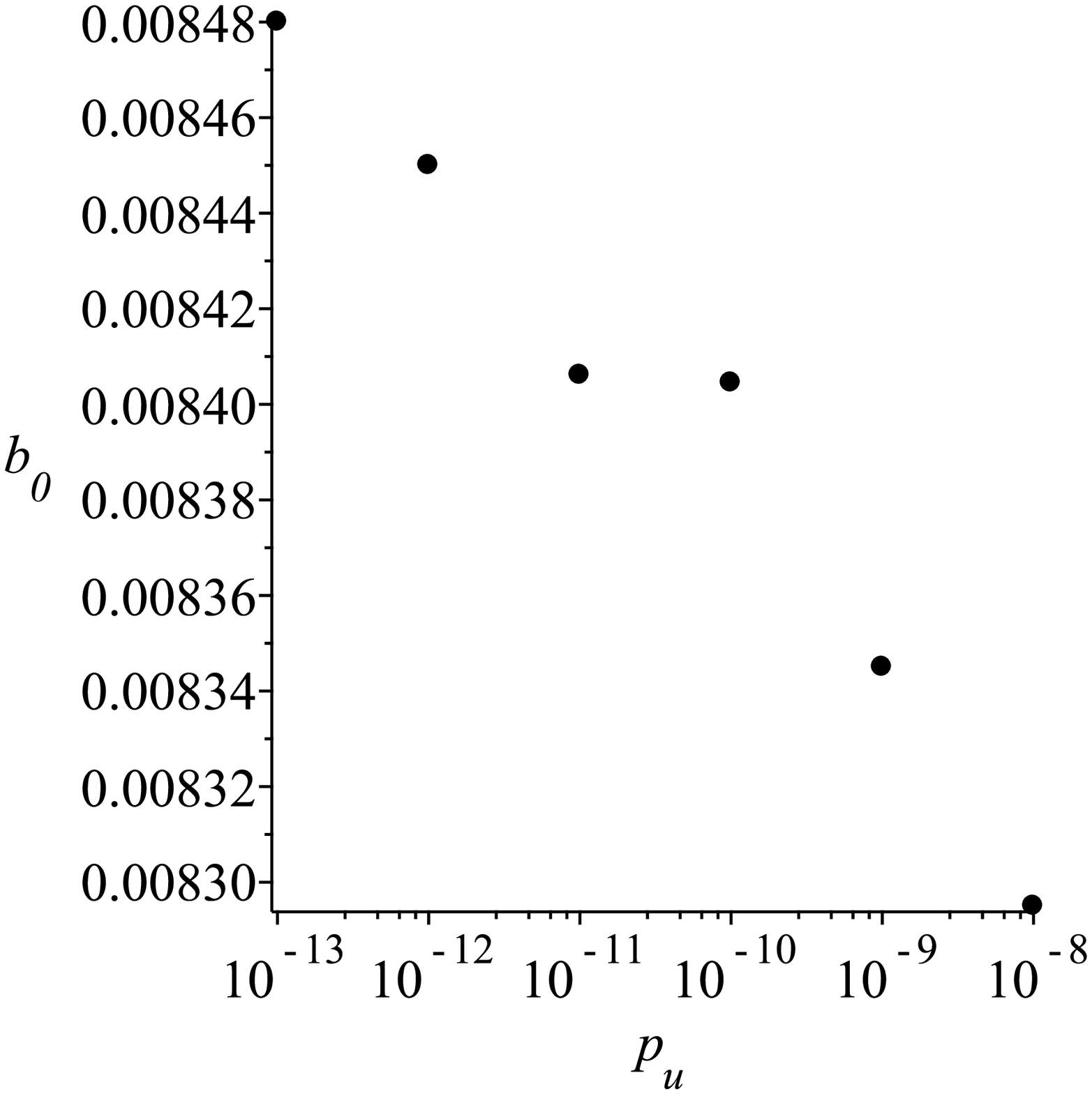}
\\
a \hskip 5cm b\hskip 5cm c
}
\caption{(\textbf{a}) The beginning of the matter dominated epoch $t_{\mathrm{SD}}$ as a function of the parameter $p_u$ for $\chi_2=1,\;M_0=-1,\;M_1=4,\;M_2=0.001, p_u=10^{-12}, \alpha=1.1$. (\textbf{b}) The $f_1$ parameters used. (\textbf{c}) The $b_0$ parameters used.}
 \label{Fig4}
 \end{figure}

Another observation is that the position of the effective potential slope is parameter dependent. 
We would like to use this freedom to set $\phi(0)=0$, which we consider ``more physical'' as well, but this is not a priority of our current numerical investigations.
Note that the example on Figure \ref{Fig1} is exactly with $\phi(0)=0$, while the data in Figures \ref{Fig2}--\ref{Fig4} is for a family of solutions for which $\phi(0)=-2.7$.

Here, we use 
$ b_0>0 $ in opposition to $b_0=-0.52$ in \cite{1408.5344} because for negative $b_0$, we have problems with $\rho$.

Thus far, we have not performed a thorough consideration of the model in the whole parameter space and even in the reduced one as described above.
However, everywhere we look, we observe only three types of deceleration-acceleration sequences in the Universe's evolution:\\
(a) only deceleration;\\
(b) deceleration followed by acceleration. 
The acceleration can be (i) very strong, i.e., we have inflation, in which case the deceleration epoch is extremely short,
or (ii) very slow, i.e., corresponding to acceleration with the cosmological constant as given in Equation (\ref{coscon}).
Which of these options is realized depends on the asymptotic value of $\phi$.\\
We also observe a (c) ``physically realistic'' Universe with four epochs: a short first deceleration epoch (FD), early inflation (EI), a second deceleration (SD), which we interpret as radiation and matter determined epochs together, and finally, an infinite slow accelerating expansion (AE). 
Some examples of this type of ``physically realistic'' evolution are shown in Figures \ref{Fig1}--\ref{Fig4}, and hereafter, we consider only this type of evolution.

The deceleration, with which any type of evolution starts, is due to the singularity of $\rho$ at $a=0$.

The type (c) evolution represents just a small part of the parameter space, very sensitive to fine-tuning.
 Thus, a significant part of the numerical work is to find the points in the parameter space for which we observe the physically realistic behavior. Numerically, this means that the second derivative of the scale factor has to cross the t-axis in three points (namely, $t_{\mathrm{EI}}$, $t_{\mathrm{SD}}$, and $t_{\mathrm{AE}}$), turning the problem into a root-finding problem. 
In order to find the zeros of the function with arbitrary precision, we use the one-dimensional Muller algorithm on the polynomial approximation of $\ddot{a}(t)$ (Equation \eqref{sys2} or Equation \eqref{dda} for independent confirmation) provided by Maple.
In the units we use,  $t_{\mathrm{SD}} \sim 10^{-50}$ and $t_{\mathrm{AE}} \sim 0.71$. 
At the moment, we are unable to reproduce $t_{\mathrm{SD}}$, but this seems to be due only to numerical problems, while there is no indication that this is impossible in principle.

The main observation in our work is that it is impossible to start the evolution from the left plateau and to obtain a physically realistic solution as defined above, no matter what is the inflaton's initial velocity. 
 If one starts from the left plateau, one ends with a Type b(i) Universe. 
 
The only way to get a physically realistic solution is to start from the slope and roll down to the right plateau. 
An example of a physically realistic solution is shown in Figure \ref{Fig1}. 
We used the parameter $b_0$ to set $t_{\mathrm{AE}}\sim0.71$ and parameter $f_1$ to ensure $a(1)=1$.

The two different regions of exponential expansion of the Universe are well visible on the logarithmic plot of $a(t)$ in Figure \ref{Fig1}, Panel (a). 

The other phases of the Universe evolution are easily traceable in Figure
 \ref{Fig1}, Panel (b), where the time dependence of $\ddot{a}(t)$ is
 plotted, and also from Panel (d), where the EOS characterizing parameter
 $w(t)=p(t)/\rho(t)$  is plotted. The asymptotics of $w(t)$ is given in
 Equation(\ref{asym2}).
 We can see that the Universe passes through the following stages: 
\begin{enumerate}
 \item At $t_0=0$ we observe the EOS of ultra-relativistic matter with $w=1/3$.
The existence of this phase does not contradict the observations, because currently, we have information solely from the time after the initial inflation;
 \item Initial inflation with EOS of dark energy $w \to -1$;
 \item Matter domination stage where $w>-1/3$;
 \item Accelerated expansion with $w<-1/3$.
\end{enumerate}

An important numerical result is that the inflaton scalar field $\phi$ tends to a constant (see Figure~\ref{Fig1}, Panel (c)), i.e., the theory predicts the existence of a scalar field with in general a nonzero average value in the late Universe.

As we have explained earlier for any set of suitable parameters, we can use $b_0$ to set $t_{\mathrm{AE}}\sim0.71$ and parameters $f_1$ to ensure $a(1)=1$; however, we have problems setting $t_{\mathrm{SD}} \sim 10^{-50}$. 
We investigated the influence of other parameters on the value of $t_{\mathrm{SD}}$.
As was expected, the parameter $\alpha$ was crucial for this value, which is illustrated in Figure \ref{Fig2}.

We can immediately draw two conclusions from Figure \ref{Fig2}:
First, a bigger $\alpha$ ensures smaller $t_{\mathrm{SD}}$. 
Thus, $\alpha\to 0$ advocated in \cite{1408.5344} seems implausible.
Second, it seems impossible to obtain realistic $t_{\mathrm{SD}}$ solely by increasing $\alpha$.

Another parameter that significantly affects the value of $t_{\mathrm{SD}}$ is $f_2$; see Figure \ref{Fig3}, from which we can make the following two conclusions:
First, smaller $f_2$ ensures smaller $t_{\mathrm{SD}}$. 
Second, it seems impossible to obtain realistic $t_{\mathrm{SD}}$ solely by decreasing $f_2$.

The third parameter on which $t_{SD}$ depends strongly is $p_u$; see Figure \ref{Fig4}. 
The results prompt that higher $p_u$ ensures smaller $t_{SD}$.

The influence of the other parameters in the model on $t_{\mathrm{SD}}$ is not so easy to track, because they are interconnected through Equation (\ref{coscon}) and because they determine the position of the effective potential slope and, thus, the possible value of $\phi(0)$. Therefore, we prefer to consider them as determining different families of solutions.

Finally, we want to 
note that starting the inflaton evolution on the slope of the effective potential can raise a question about maintaining small slow-roll parameters during inflation.
However, there are two mechanisms in the model preventing the appearance of large kinetic energy. One has to note that since the inflaton equation is not in the standard form, the friction should not be considered as a dissipative process, but as an exchange of energy between the two scalar fields. 

Second, depending on the parameter choice, the inflaton can perform rather peculiar movement; see Figure \ref{Fig1}c. 
In this case, the inflaton starts its evolution with $\phi(0)=0$ and $\dot{\phi}(0)=0$.
Its position corresponds almost exactly to the middle of the effective potential slope with $d U_{eff}/d \phi <0$.
Nevertheless, the initial movement of the inflaton is towards negative values, i.e., upwards of the potential slope, a situation resembling the stability of the L4 and L5 Lagrange points.
As a result, for a relatively long period of time, $U_{eff}(\phi(t))$ does not change significantly, and the inflaton kinetic energy remains small. 
 These effects together with the small value of $p_u$ ensure almost pure exponential growth of $a(t)$.
This in turns gives that the slow-roll parameter $\epsilon=-\dot{H}/H^2$ during inflation is $\epsilon << 1$, because $\epsilon$ effectively measures the ``exponentiality'' of $a(t)$. 

\section{Conclusions}
We studied numerically the inflaton model of Guendelman--Nissimov--Pacheva by integrating the coupled differential system governing the inflaton fields and metric scaling parameter. 
Our calculations confirmed that qualitatively, this model can successfully describe the evolution of the Universe, since it naturally reproduces its main phases.
Quantitatively, the emerging picture seems different than the one sketched in \cite{1408.5344,1609.06915}.
This requires some additional work, both theoretical and numerical, in order to clarify the situation.

We showed that the inflationary region of the effective potential is its slope, while the evolution starting from the higher plateau of the step-like effective potential is nonphysical. Our numerical work showed that the type of evolution of the Universe, i.e., the number of stages one observes, depends on the starting position on the effective potential, while the other parameters define the shape of the potential and the time scales of the different stages. A connecting result is that the inflaton field tends to a constant, thus predicting the existence of a non-zero averaged scalar field in the current Universe. The value of this constant depends mainly on the initial value of the inflaton field.

We studied the dependence on the parameters of the time at which the inflation ends. We demonstrated that it is possible to obtain shorter inflation by making the
effective potential higher, steeper, and if there is more dark matter.

\vspace{6pt} 



\authorcontributions{Conceptualization, D.S. and M.S.; methodology, D.S. and M.S.; analysis, D.S. and M.S.; writing, original draft preparation, D.S. and M.S.; writing, review and editing, D.S. and M.S.}

\funding{
The work is supported by the Bulgarian NSF grant DN-18/1/10.12.2017 and by the Bulgarian NSF grant 8-17. D.S. is also partially supported by COST Actions CA18108.}

\acknowledgments{
It is a pleasure to thank Emil Nissimov and Svetlana Pacheva for the discussions. }

\conflictsofinterest{{The authors declare no conflict of interest}.} 

%


\reftitle{References}


\begin{thebibliography}{999}
 
\bibitem{cosmo0} Marochnik, L. {Dark Energy and Inflation in a Gravitational Wave Dominated Universe}. \emph{Gravitat. Cosmol.} \textbf{2016}, {\bf 22}, 10--19. 
 
\bibitem{Linde} Linde, A. {Inflationary Cosmology after Planck}.  \emph{arXiv} \textbf{2013},  arXiv:1402.0526.

\bibitem{Debono} Debono, I.; Smoot, G.F. {General Relativity and Cosmology: Unsolved Questions and Future Directions}.  \emph{Universe} \textbf{2016},  \emph{2(4)}, 23. 

\bibitem{Oda} Oda, I. {Classical Weyl transverse gravity}.  \emph{Eur. Phys. J. C} \textbf{2017},  \emph{77}, 284. 

\bibitem{Bars} Bars, I.; Steinhardt, P.; Turok, N. {Local conformal symmetry in physics and cosmology}. \emph{Phys. Rev. D} \textbf{2019}, {\emph{89}},~043515. 

\bibitem{1904.04493} Tang, Y.; Wu, Y. {Weyl Symmetry Inspired Inflation and Dark Matter}.  \emph{arXiv}  \textbf{2019},  arXiv:1904.04493. 

\bibitem{Edery} Edery, A.; Nakayama, Y. {Palatini formulation of pure R2 gravity yields Einstein gravity with no massless scalar}.  \emph{Phys. Rev. D} \textbf{2019}, \emph{99}, 124018.

\bibitem{Planck2018} Planck Collaboration.  {Planck2018 results. X. Constraints on inflation.}    \emph{arXiv} \textbf{2019},  arXiv:1807.06211.

\bibitem{1807.02376} Rubio, J. {Higgs inflation}.  \emph{Phys. Rev. D} \textbf{2019}, \emph{99}, 124018.


\bibitem{ref01} Guendelman, E. {Scale invariance and vacuum energy}. \emph{Mod. Phys. Lett. A} \textbf{1999}, {\emph{14}}, 1043--1052.

\bibitem{ref01_1} Guendelman, E.; Kaganovich, A. {Dynamical measure and field theory models free of the cosmological constant problem} \emph{Phys. Rev. D} \textbf{1999}, {\emph{60}}, 065004.

\bibitem{ref01_2} Guendelman, E.; Katz, O., {Inflation and transition to a slowly accelerating phase from SSB of scale invariance.} \emph{Class.} \emph{Quantum Grav.} \textbf{2003}, {\emph{20}}, 1715--1728. 

\bibitem{ref01_31}Guendelman, E.I.; Labrana, P.  {Connecting The Non-Singular Origin of the Universe, The Vacuum Structure and The Cosmological Constant Problem}. \emph{Int. J. Mod. Phys. D} \textbf{2013}, \emph{22},  1330018.
\bibitem{ref01_32}Guendelman, E.I.; Singleton, D.; Yongram, N. {A two measure model of dark energy and dark matter}. JCAP \textbf{2012}, \emph{1211},  044. 

\bibitem{ref01_33}Guendelman, E.I.; Nishino, H.; Rajpoot, S. {Scale Symmetry Breaking From Total Derivative Densities and the Cosmological Constant Problem}. \emph{Phys. Lett. B} \textbf{2014}, \emph{732},  156--160.  

\bibitem{1407.6281} Guendelman, E.I.; Nissimov, E.; Pacheva, S. {\it Unification of Inflation and Dark Energy from Spontaneous Breaking of Scale Invariance}; 
 Dragovic, B., Salom, I.,  Eds.; Belgrade Inst. Phys. Press: Belgrade,  Serbia,  2015;  pp. 93--103.
%
\bibitem{1408.5344}Guendelman, E.; Herrera, R.; Labrana, P.; Nissimov, E.; Pacheva, S. {Emergent Cosmology, Inflation and Dark Energy}.  \emph{Gen. Relativ.  Gravit.} \textbf{2015}, {\emph{47}}, 10.  
   
%
\bibitem{1609.06915} Guendelman, E.; Nissimov, E.; Pacheva, S. {Quintessential Inflation, Unified Dark Energy and Dark Matter, and Higgs Mechanism}. \emph{Bulg. J.  Phys.}  \textbf{2017}, \emph{44}, 
 15--30.
%
\bibitem{1507.08878}Guendelman, E.; Herrera, R.; Labrana, P.; Nissimov, E.; Pacheva, S. {Stable Emergent Universe---A Creation without Big-Bang}.  \emph{Astronomische Nachr.} \textbf{2015}, {\emph{336}},  810--814.
 
%
\bibitem{1603.06231} Guendelman, E.; Nissimov, E.; Pacheva, S. {Gravity-Assisted Emergent Higgs Mechanism in the Post-Inflationary Epoch}.  \emph{Int. J.  Modern Phys. D}   \textbf{2016},
\emph{25}, 
 1644008.
%
\bibitem{1610.08368} Staicova, D.; Stoilov, M. {Cosmological Aspects Of A Unified Dark Energy And Dust Dark Matter Model}.  \emph{Mod. Phys. Lett. A}  \textbf{2017},  \emph{ 32},  1750006. 
%
\bibitem{1801.07133} Staicova, D.; Stoilov, M. {Cosmological solutions from models with unified dark energy and dark matter and with inflaton field}. Springer Proc.Math.Stat. \textbf{2017}, \emph{255}, 251--260.
 
\bibitem{1906.08516} Staicova, D.; Stoilov, M. {Cosmology from multi-measure multifield model}. \emph{ Int.  J.  Mod. Phys. A} \textbf{2019},   \emph{34},~1950099.






\end{thebibliography}
\end{document}